\documentstyle[12pt,epsf]{article}
\newcommand{\wt}{\widetilde}

\newcommand{\be}{\begin{equation}}
\newcommand{\ee}{\end{equation}}
\newcommand{\ben}{\begin{eqnarray}\displaystyle}
\newcommand{\een}{\end{eqnarray}}
\newcommand{\refb}[1]{(\ref{#1})}
\newcommand{\p}{\partial}

\begin{document}

{}~ \hfill\vbox{\hbox{hep-th/9805170}\hbox{MRI-PHY/P980551}
}\break

\vskip 3.5cm

\centerline{\large \bf Tachyon Condensation on the Brane
Antibrane System}

\vspace*{6.0ex}

\centerline{\large \rm Ashoke Sen
\footnote{E-mail: sen@mri.ernet.in}}

\vspace*{1.5ex}

\centerline{\large \it Mehta Research Institute of Mathematics}
 \centerline{\large \it and Mathematical Physics}

\centerline{\large \it  Chhatnag Road, Jhoosi,
Allahabad 211019, INDIA}

\vspace*{4.5ex}

\centerline {\bf Abstract}

A coincident D-brane - anti-D-brane pair has a tachyonic
mode. We present an argument showing that at the classical minimum
of the tachyonic potential the negative energy density associated
with the potential exactly cancels the sum of the tension of the
brane and the anti-brane, thereby giving a configuration of zero
energy density and restoring space-time supersymmetry.

\vfill \eject

\baselineskip=18pt

It has been known for sometime that a coincident D-brane $-$
anti-D-brane pair contains a tachyonic
excitation\cite{GREEN,BANSUS,GRGUP,PERI}. In a 
previous paper\cite{NB}
it was conjectured that at the minimum of the tachyon potential,
the negative contribution to the energy density from
the potential exactly cancels the sum of the tensions of the
brane and the anti-brane, thereby giving a configuration of zero
energy density (and hence restoring space-time
supersymmetry). Although this proposal might sound radical, there
are many known examples where tachyon condensation restores
space-time supersymmetry\cite{POLSTR,NARA,NARAT}, and the present
example is just an extension of these earlier examples. 
In this paper we shall offer a `proof' of this conjecture.
Although this will not be a rigorous mathematical proof; at
least it should serve the purpose of putting this conjecture on a
firmer footing.

The strategy that we shall adopt will follow closely that of
\cite{NARA}; and indeed most of the results we shall require are
already contained there. 
(For earlier references related to this subject, see \cite{EARLY}.)
For definiteness we shall focus on the
Dirichlet 2-brane, but the extension to other branes should be
straightforward. We take the coincident 2-brane 
anti-2-brane system, and wrap it on a two dimensional torus $T^2=S^1\times
S^1$, of radii $R_1$ and $R_2$ respectively. (For simplicity we
shall set $\alpha'=1$ and measure all lengths and masses in
string units.) We now introduce one unit of magnetic flux on both
the brane and the anti-brane; and the signs of the magnetic
fields
are chosen such that each corresponds to $+1$ unit of D0-brane
charge. The strength of the magnetic field on each brane is given
by:
\be \label{e1}
F_{12} = {2\pi\over V}\, ,
\ee
where $V=4\pi^2 R_1 R_2$ is the area of the torus. In
the limit $R_1\to\infty$, $R_2\to\infty$, the magnetic field
strength goes to zero and we expect to 
recover the physics of the original brane $-$
anti-brane system. 

We shall however begin by analyzing this system in the 
$R_1,R_2\to 0$ limit. In this limit the system is best described
by going to the T-dual version, in which we make $R_i\to (1/R_i)$
duality transformation on both the circles. This gives a dual
torus $\wt T^2$ with radii
\be \label{e2}
\wt R_i = {1\over R_i}\, , \qquad i=1,2\, ,
\ee
and coupling constant
\be \label{eg1}
\wt g = {g\over R_1 R_2}\, ,
\ee
where $g$ is the coupling constant of the original
theory.\footnote{During the process of varying the parameters
$R_1$ and $R_2$ to take various limits, we shall keep both $g$
and $\wt g$ small, so that the classical description is good in
the original theory as well as in the dual theory.}
Under this duality transformation the 0-brane and the 2-brane
charges get interchanged. Thus
the wrapped D2-brane with one unit of magnetic flux gets converted
to a wrapped D2-brane with $1$ unit of magnetic flux, whereas the
wrapped anti-D2-brane with one unit of magnetic flux gets
converted to a wrapped D2-brane with $-1$ unit of magnetic flux.
When $\wt R_1$ and $\wt R_2$ are large, this system may be
described by an effective supersymmetric U(2) gauge theory on this
dual torus\cite{WITT} 
in the presence of a background gauge field of the form:
\be \label{e3}
A_1=0, \qquad 
A_2 = {2\pi x^1 \sigma_3\over \wt V}\, ,
\ee
where $x^1,x^2$ denote the directions of the two circles of
$\wt T^2$, $A_\mu$
denotes the component of the U(2) gauge field along the $\mu$th
circle,
$\sigma_i$ are the Pauli matrices, and $\wt V=4\pi^2\wt R_1\wt
R_2$ is the volume of the dual torus $\wt T^2$. This gauge field $A$
satisfies the boundary conditions
\ben \label{e4}
A(x^1=2\pi \wt R_1,x^2) &=& \Omega_1\circ A(x^1=0,x^2)\, , \nonumber \\
A(x^1,x^2=2\pi \wt R_2) &=& \Omega_2\circ A(x^1,x^2=0)\, , 
\een
where $\Omega_\mu$ are gauge transformation parameters:
\be \label{e5}
\Omega_1 = \exp(i x^2 \sigma_3/\wt R_2)\, , \qquad \Omega_2 = 1\, ,
\ee
and $\Omega_\mu\circ A$ denotes the gauge transform of $A$ by
$\Omega_\mu$. $\Omega_1$ and $\Omega_2$ satisfy the relation:
\be \label{e6}
\Omega_2(x^1=2\pi \wt R_1) \Omega_1(x^2=0) = 
\Omega_1(x^2=2\pi \wt R_2) \Omega_2(x^1=0)\, .
\ee

{}From eq.\refb{e3} we see that the gauge field configuration
lies fully inside the SU(2) part of the gauge group, and does
not have any U(1) component. If we now consider fluctuations of
the SU(2) gauge fields around the background gauge field
configuration given in \refb{e3}, satisfying the boundary
conditions given in \refb{e4}, \refb{e5}, we shall find 
tachyonic modes\cite{NARA}. 
It is easy to verify that the mass spectrum of the tachyonic
modes is
identical to that calculated directly from the spectrum of
the original string theory
before duality transformation. In the gauge theory the presence
of these tachyonic modes simply reflect the fact that the gauge
field configuration given in \refb{e3}, which gives rise to a
non-vanishing field strength and hence a
positive definite energy density on the brane, does not correspond to the
minimum energy configuration subject to the boundary conditions
\refb{e4}, \refb{e5}. To see this let us note that with the help
of a gauge transformation $g(x^1,x^2)$ on $A_\mu(x)$, 
we can make both $\Omega_1$ and
$\Omega_2$ to be
identity, since, as can be seen from \refb{e6}, there is no
obstruction to this choice\cite{HOOF}. In particular, we can
choose $g(x^1,x^2)$ such that 
\ben \label{enew1}
&& g(x^1=0,x^2)= \exp(ix^2\sigma_3/\wt R_2)\, , \quad
g(x^1=2\pi \wt R_1, x^2) = 1\, , \nonumber \\
&& g(x^1, x^2=2\pi \wt R_2)=g(x^1,x^2=0)\, , \quad
{\p\over \p x^1}g(x^1,x^2)=0\,
\, \hbox{at} \,\, x^1=0, 2\pi \wt R_1\, .
\een
The existence of a $g(x^1,x^2)$ satisfying these conditions can
be proved by noting that $g(x^1=0,x^2)$ describes a map from
$S^1$ to the SU(2) group manifold. Since SU(2) is simply connected,
it is possible to find an interpolating map $g(x^1,x^2)$ such
that $g(x^1=2\pi \wt R_1, x^2)$ is the identity element of the group.
The transformed field $g\circ A_\mu$  satisfies
periodic boundary condition along $x^1$ and 
$x^2$, and thus the lowest
energy configuration corresponds to $g\circ A_\mu=0$ and hence
$g\circ F_{\mu\nu}=0$. We can go back to 
the original gauge
in which $\Omega_\mu$ are given by \refb{e4} with the help of a
reverse gauge transformation $g^{-1}(x^1,x^2)$. 
This will give rise to a
non-trivial gauge field configuration, but the field strength
$F_{\mu\nu}$ will continue to vanish.

This shows that the classical minimum energy configuration after
`tachyon condensation' corresponds to zero field strength for
both the U(1) and the SU(2) gauge fields, and hence the energy
per unit area will be given by the sum of the tensions of the two
D2-branes. Since in the unit we are using each $D$-brane has
tension $1/(4\pi^2 \wt g)$, $\wt g$ being the string coupling constant,
and the total area of each brane is $4\pi^2 \wt R_1 \wt R_2$,
we see that the total mass of this system is given by:
\be \label{e7}
M = {2\wt R_1 \wt R_2\over \wt g}\, .
\ee
Note that this saturates the BPS bound for a pair of D2-branes
wrapped on the dual torus $\wt T^2$. It is also possible to 
argue that once
quantum fluctuations are taken into account, there is a
quantum ground state of the system with exactly
the same mass\cite{SEMA,SETSTE}.\footnote{These states are needed
for U-duality, as they are related to the Kaluza-Klein modes
carrying two units of momentum along the internal directions of
the torus.} 

Since the classical ground state of the system saturates BPS
bound, we would expect that it would continue to saturate BPS
bound as we change the parameters of the theory continuously, and
hence its mass at the classical minimum will continue to be given
by \refb{e7}. Let
us now express the mass formula \refb{e7} in terms of the
original variables $R_1$, $R_2$ and $g$. Since
the string metric does not change under a T-duality
transformation, eq.\refb{e7}, when expressed in
terms of the original variables using eqs.\refb{e2}, \refb{eg1}, 
takes the form:
\be \label{e8}
M = {2\over  g}\, .
\ee
Now recall that the original system was a D2-brane and an
anti-D2-brane wrapped on a torus $T^2$ of 
area $4\pi^2 R_1 R_2$, with one
unit of magnetic flux on each of the branes. Thus from \refb{e8}
we see that at the minimum of the tachyon potential, the energy
per unit area on the  brane $-$ anti-brane system is given by:
\be \label{e9}
{M\over 4\pi^2 R_1 R_2} = {1\over 2\pi^2 R_1 R_2 g}\, .
\ee
If we now take the limit
$R_1,R_2\to\infty$, the energy per unit area 
goes to zero. On the other hand, as has already been
argued before, in this limit the magnetic field strength on the
brane and the anti-brane goes to zero and we expect to recover
the physics of the brane $-$ anti-brane system without any
magnetic flux. Thus we deduce that at the classical
minimum of the potential, the energy per unit area of the brane
$-$ anti-brane system vanishes.

This concludes our `proof'. If we assume that a similar result
holds for the D-string anti-D-string pair in type I string
theory, then in the spirit of \cite{NB} we can identify the
stable non-BPS
SO(32) spinor states in type I theory\cite{NONBP}
as the `tachyonic soliton'
on the D-string - anti-D-string pair in which the tachyon field
changes from $-T_0$ to $+T_0$ as we go from far left to the far
right on the string. Here $\pm T_0$ denote the locations of the
minima of the tachyonic potential at which the tension of the
D-string  anti-D-string pair is canceled by the tachyon
potential. In order to see that this represents an SO(32) spinor
state, we can compactify type I theory on a circle of large
radius, and consider a D-string anti-D-string pair wound around
that circle. If one of them carries a non-trivial $Z_2$ Wilson
line\cite{POLWIT} and the other carries trivial $Z_2$ Wilson
line, then the combined system represents an SO(32) spinor, being
a combination of a spinor and a singlet state. In this case the
tachyon associated with the open string stretched between the
D-string and the anti-D-string must be anti-periodic as we go
around the circle. The kink solution described above precisely
satisfies such a boundary condition. Bound states of wound
D-strings carrying spinorial and winding charges have been
discussed recently in \cite{NAREC}.

\end{document}